# Topological phase transition in Dirac fermionic heterostructures

Jeongwoo Kim[1], Jinwoong Kim[1], Ki-Seok Kim[1,3] & Seung-Hoon Jhi[1,2*]

[1]*Department of Physics, Pohang University of Science and Technology, Pohang 790-784, Republic of Korea*

[2]*Division of Advanced Materials Science, Pohang University of Science and Technology, Pohang 790-784, Republic of Korea*

[3]*Asia Pacific Centre for Theoretical Physics, Pohang University of Science and Technology, Pohang 790-784, Republic of Korea*

Abstract

Materials with non-trivial topology in their electronic structures enforce the existence of helical Dirac fermionic surface states. We discovered emergent topological phases in the stacked structures of topological insulator and band insulator layers where the surface Dirac fermions interact to each other with particular helicity ordering. Using first-principles calculations and a model Lagrangian, we explicitly demonstrated that such helicity ordering occurs in real materials of ternary chalcogen compounds and determines their topological insulating phase. Our results reveal the rich collective nature of interacting surface Dirac fermions, and pave the way for utilizing topological phases for technological devices such as non-volatile memories.

Topology characterizes the invariance of objects under continuous deformation. Operations of non-continuous deformation like gluing can make a new manifold from two distinguishable topological objects. Their manifestation in physical properties of real materials, however, has rarely been discussed. Topological insulators, which possess non-trivial topology in the electronic structure with an energy gap, have helical Dirac fermions at surfaces dictated by the time-reversal symmetry [1-4]. Stacking of topological insulators (TIs) alternated by band insulators (BIs) provides an interesting platform to explore emerging phenomena associated with topological phases in solids. For example, hybrid structures of different topological matters are proposed to detect hypothetical particles such as magnetic monopole, the axion, and the Majorana fermion [5-7]. Helical Dirac fermions have linear energy-momentum relation that forms a cone (Dirac cone) at a particular momentum, and they are robust against non-magnetic perturbations [8-10]. Alternating arrays of TI and BI layers thus generate the superlattice of Dirac fermions. When the helical surface states are close to each other – for instance, in thin TI film (slab hereafter) or when two TI slabs are in close contact, the Dirac fermions interact with each other and come to have a mass gap [11, 12]. The Dirac fermions have two types of helicity as like neutrinos, say right-handed and left-handed, and the interactions depend on them. Similar to the magnetization ordering in magnetic superlattices [13, 14], different helicity orderings can occur in the Dirac fermionic superlattice and should be considered for understanding the collective behavior of Dirac fermions.

Ternary chalcogen compounds of Ge-Sb-Te, Ge-Bi-Te, and Ge-Bi-Se, which are also well-known structural phase-change materials used for non-volatile memory devices [15-17], have a layered structure with stacking of Ge-Te and Sb-Te (Bi-Te or Bi-Se). Interestingly, they turn out to be topological insulator or band insulator depending on their composition and stacking sequence [18, 19]. Figure 1(a) shows the atomic structure of $Ge_2Bi_2Te_5$ (GBT225) and $Ge_2Sb_2Te_5$ (GST225) considered in this study. The $Z_2$ index that defines the topological phase of materials [3, 20] and the parity check in their band structures show that the two compounds have different topological phases. Calculated surface band structures of GST225 and GBT225 in Fig. 1(b) and (c) confirm their topological phase. GBT225 has a clear Dirac cone but GST225 has gapped surface states. The plot of squared wave-functions of the states near the Fermi level in Fig. 1(d) contrasts the feature of the surface states in GST225 and GBT225 slab structures. For GST225, the states are mostly localized at each Te-Te layer while retaining some characters of the surface states of $Sb_2Te_3$. For GBT225, the states are mostly localized near the surface and have truly the helical structure of topological insulator. The low energy states near the Fermi level in these compounds highlight the intriguing aspect of interacting Dirac fermions. In this Letter, we studied topological phase transition in ternary chalcogen compounds using a model Lagrangian and first-principles methods. We particularly investigated the helicity ordering of the surface states from TI layers and its role for emerging topological phases.

Calculations were conducted using first-principles self-consistent pseudopotential method [21] as implemented in the Vienna *ab initio* simulation package (VASP) [22]. The exchange-correlation of electrons was treated within the generalized gradient approximation (GGA) in the form of Perdew-Burke-Ernzwehof (PBE)[23], and the spin-orbit coupling (SOC) was included in the self-consistent calculations. The projector-augmented wave (PAW) potentials [24] as supplied in the VASP package were used for atomic potentials. We considered a series of ternary chalcogen compounds, $Ge_l Sb_m Te_n$ (GST*lmn*), $Ge_l Bi_m Te_n$ (GBT*lmn*), and $Ge_l Bi_m Se_n$ (GBS*lmn*) ($l$=1,2,4; $m$=2,4,6,8; $n$=4,5,6,8,11,14) to investigate their topological insulating phase. These compounds can be considered as a short-period superlattice of band insulating layer (GeTe or GeSe) and topological insulating layer ($Sb_2Te_3$, $Bi_2Te_3$ or $Bi_2Se_3$) because low energy states near the Fermi level are hardly affected by the hybridization of TI and BI layers. Figure 2 shows the calculated band gap and the relative thickness of GeTe and $Sb_2Te_3$ layers in GST compounds along with their topological insulating phases. The energy gap occurs at $\Gamma$ point and originates from $Sb_2Te_3$ layers. We observe an interesting transition from band insulating phase into topological insulating phase as the thickness of TI layers relative to that of BI layers is increased. This transitional behavior indicates that the topological phase transition can occur depending on certain materials properties related to the thickness of TI and BI layers.

In order to understand the topological phase in the stacking structures, we consider a superlattice of Dirac fermions developed at interfaces between the layers of different $Z_2$ indices as shown in Fig. 3(a). The Dirac fermions in the superlattice have both intralayer and interlayer interactions. The Dirac fermionic superlattice is characterized by three parameters as schematically shown in Fig. 3(b); the intrinsic mass gap ($\Delta$) due to the intralayer interaction (the interaction between the surface states in the same TI layer), the coupling strength ($V$) between Dirac fermions and band electrons of the BI layers, and the band gap ($\varepsilon_c$) of BI layers. These parameters ($\Delta$, $V$, and $\varepsilon_c$) will determine the topological phase of the Dirac fermionic superlattice. Here we are left with a degree of freedom of how to order the helicity of the Dirac fermions. Basically, two possible choices of the helicity ordering are available between two Dirac fermions in the adjacent TI layers at z=0 and z=$L$; the same helicity (co-helicity) or the opposite helicity (counter-helicity). Figure 3(c) shows the schematic diagram of the helicity ordering in two adjacent layers. Since the Dirac fermions can interact with each other of the same helicity, the resulting topological phase of interacting Dirac fermions should depend on the helicity ordering. This is similar to joining two Möbius strips. The mass gap acts as scissors to cut the strip and the coupling $V$ (or effective coupling) as glue to link the strips. We may have a Möbius strip or a strip with double twisting depending on pairing the ends cut by scissors. From topological aspect, the single Möbius represents the topological-insulating phase and the doubly-twisted strip corresponds to the non-topological-insulating phase.

The effective Lagrangian for the Dirac fermions with the intralayer and interlayer interactions can be written as

$$\mathcal{L}_{eff} = \psi^{\dagger}_{\vec{k}\alpha}(0)[-i\omega + v\vec{k}\cdot\vec{\sigma}_{\alpha\beta} + \Delta\sigma^{z}_{\alpha\beta}]\psi_{\vec{k}\beta}(0) + \psi^{\dagger}_{\vec{k}\alpha}(L)[-i\omega \pm v\vec{k}\cdot\vec{\sigma}_{\alpha\beta} \pm \Delta\sigma^{z}_{\alpha\beta}]\psi_{\vec{k}\beta}(L)$$
$$+ [\psi^{\dagger}_{\vec{k}\alpha}(0) + \psi^{\dagger}_{\vec{k}\alpha}(L)]G_0[\psi_{\vec{k}\alpha}(0) + \psi_{\vec{k}\alpha}(L)]$$

where $\psi_{\vec{k}\alpha}(0)$ and $\psi_{\vec{k}\alpha}(L)$ are the fields of the surface Dirac fermions with the Fermi velocity $v$ at $z = 0$ and $z = L$, respectively, with $\vec{k}$ representing their momentum in the $xy$ plane and $\alpha,\beta$ the spin indices. This Lagrangian is valid at a low energy limit where the bulk gap of TI ($\varepsilon_g$) is much larger than the mass gap $\Delta$ so that the bulk states are decoupled from the surface states. $\vec{\sigma} = (\sigma_x, \sigma_y, \sigma_z)$ are the Paulic matrices and $G_0$ is the Green's function of band insulating layers, $G_0 = \sum_{n=-\infty}^{\infty} V^2 (i\omega + \varepsilon_c - \frac{2\pi^2 n^2}{mL^2} - \frac{\vec{k}^2}{2m})^{-1}$, within the periodic boundary condition. The sign in the second term represents the helicity ordering. We note that at a level of the second order perturbation, the co-helicity is favored in the "resonant condition" $\varepsilon_g \sim 2\Delta + \varepsilon_c$. When the energy gap of band insulator is increased, the energy gain by hybridization decreases as $(2\Delta + \varepsilon_c)^{-1}$ for the co-helicity and as $(2\Delta + \varepsilon_c)^{-1/2}$ for the counter-helicity, favoring the counter-helicity. We solved the secular equation from the Lagrangian to find the dispersion relation $\omega(\vec{k})$ for the Dirac fermions and to determine the band inversion as a function of $V$, $\Delta$ and $\varepsilon_c$. We found that the topological insulating phase emerges in the Dirac fermionic superlattice only for the co-helicity. When Dirac fermions in the adjacent TI layers are in the co-helicity, the effective interlayer interaction pushes upward in energy the upper branch of Dirac cone at $z = 0$ and downward the upper branch of Dirac cone at $z = L$ [right bottom panel of Fig. 3(c)]. Also, the interaction moves the lower branch of Dirac cone at $z =$ 0 upward in energy and the lower branch of Dirac cone at z = L downward. As a result, the band inversion can occur between the upper Dirac cone at z = L and the lower Dirac cone at z = 0 when the coupling strength is increased. However, for the counter-helicity, the coupling is allowed between the upper (lower) branch of Dirac cone at z = 0 and the lower (upper) branch of Dirac cone at z = L [right top panel of Fig. 3(c)]. Hence, such hybridization only leads to the widening of band gap without the band inversion.

The resulting phase diagram is shown in Fig. 4(a) in terms of $\Delta$, $V$, and $\varepsilon_c$. The red-colored region corresponds to the topological insulating phase, for which the band inversion occurs between two Dirac fermions at z=0 and L. The phase diagram is also plotted in terms of the reduced coupling constant $\tilde{V}(=V/\varepsilon_c)$ and mass gap $\tilde{\Delta}(=\Delta/\varepsilon_c)$ in Fig. 4(b) after scaling the parameters $V$ and $\Delta$ by $mL^2\varepsilon_c$ ($m$ being the electron mass) in the model Lagrangian. It is apparent that the topological insulating phase is developed when the mass gap is decreased and/or the coupling strength is increased. The band inversion occurs if the coupling strength $V$ is large enough for the Dirac fermions at adjacent TI layers to overcome the mass gap and switch the order in energy. And then the superlattice turns into TI. We carried out first-principles calculations of a series of ternary chalcogen compounds to determine their Z2 number. We also repeated the calculations for slab structures to approximate the BI and TI layers in the superlattice and obtain the mass gap and the chemical potential. Calculated band gaps of the slabs were then matched to the mass gap $\Delta$ (for TI slab) and the chemical potential $\varepsilon_c$ (for BI layer). The coupling constant $V$ is estimated by assuming that the inversion gap (at the $\Gamma$ point) in the bulk corresponds to the level splitting in the effective Lagrangian for the Dirac fermions. Using the parameters obtained from first-principles calculations, we mapped the data points $\tilde{V}$ and $\tilde{\Delta}$ of ternary chalcogen compounds in the phase diagram [Fig. 4(b)]. We note that the mass gap becomes very small when the thickness of TI layer is large, for example, more than 5QL for $Sb_2Te_3$. In such cases of very thick TI layers, the band inversion or band repulsion by *V* cannot resolve the surface states in the slab from the interface states in the bulk.

Now a question of fundamental significance is whether the helical ordering occurs in real materials and is responsible for the emergence of their topological insulating phase. We investigated the helicity ordering in two representative compounds, band insulating GST326 and topological insulating GST248. Figure 4(c) and (d) show calculated results of the spin configuration for the two compounds near the Γ point, where the Dirac cone is formed, by projecting the bands near the Fermi level into the real space. We observe clearly the counter-helicity ordering for GST326 and the co-helicity ordering for GST248 in the adjacent TI layers. The 326-compounds all have the non-topological insulating phase, and this is related to their counter-helicity ordering. Such helicity variation can be detected in the step surfaces of the compounds by utilizing circular dichroism [25]. Our results demonstrate that real materials with TI-BI stacking structure can possess two types of the helicity ordering and that their topological phase depends on it. One caution here is that the helicity mixing as in the neutrino oscillation problem can occur in the surface states but it is not considered in our model Lagrangian. The mass gap in effect allows the helicity mixing for the surface states in the TI layers, and the requirement of the co-helicity for emerging TI phases cannot be strictly enforced in such cases. In our first-principles calculations, GBT225 and GBS225 have a single TI layer in the unit cell having the counter-helicity by construction. Their TI phase is thus attributed to the band inversion by the helicity mixing. Also, putting together here, the degree of freedom in the helicity ordering assumed in our model Lagrangian disappears in the limit of $\varepsilon_c \to \infty$ or $V \to 0$.

Similar to magnetic ordering of ferro- or antiferro-magnetism, the co- and counter-helicity of Dirac fermions is crucial for the collective behavior of Dirac fermions in the lattice.

Since the mass gap, the band gap, and the coupling between band electrons and Dirac fermions depend on various external parameters such as temperature and pressure, the emergence of topological phases can exhibit rich temperature or pressure dependence. The electrical switching property of ternary chalcogen compounds can be affected significantly by their topological phase since the electronic bands originating from the helical interface states provide electrical conducting channels robust against atomic perturbations. Recent experiments that show the device performance of GST compounds depending on magnetic fields and layer stacking [26, 27] cast interesting implications in this respect. We also expect significant Joule heating near interfaces between TI and BI layers [28], which can be related to thermally-induced crystal-amorphous structural transitions of chalcogen compounds.

In summary, we studied the topological insulating phase transition in ternary chalcogen compounds. The phase diagram of topological insulating order was produced based on the superlattice model of topological insulator and band insulator layers. We mapped ternary chalcogen compounds in the phase diagram using the materials parameters obtained from first-principles calculations. We showed that topological insulating phases emerge when Dirac fermions in the superlattice interact with a particular helicity ordering, and explicitly demonstrated that two types of helicity ordering truly occur in the ternary chalcogenides and determine their topological phases. Our results can be utilized to explore the collective behavior of interacting Dirac fermions and its consequences for electronic and thermal properties of chalcogen compounds.

**Acknowledgements**

We thank Jun-Seong Kim and Sang W. Cheong for discussion and comments. This works was supported by the National Research Foundation of Korea (NRF) grant funded by the Korea government (MEST) (SRC program No. 2011-0030046 and WCU program No. R31-

2008-000-10059-0). The authors would like to acknowledge the support from KISTI supercomputing center through the strategic support program for the supercomputing application research (No. KSC-2010-C2-0008).

**Figures & captions**

Figure 1 (Color online) (a) Atomic structure of GBT225 (or GST225) in the hexagonal structure with blue balls representing Te, yellow balls Ge, and green balls Bi (or Sb). (b) and (c) Calculated surface band structures of GBT225 and GST225, respectively. The shaded regions are the bulk bands projected into surface Brillouin zone. The Fermi level is set at zero energy denoted by the dashed line. (d) The plot of squared wave functions of the states near the Fermi level for GST225 (left panel) and GBT225 (right panel). Bright (yellow green) spots represent high-density regions.

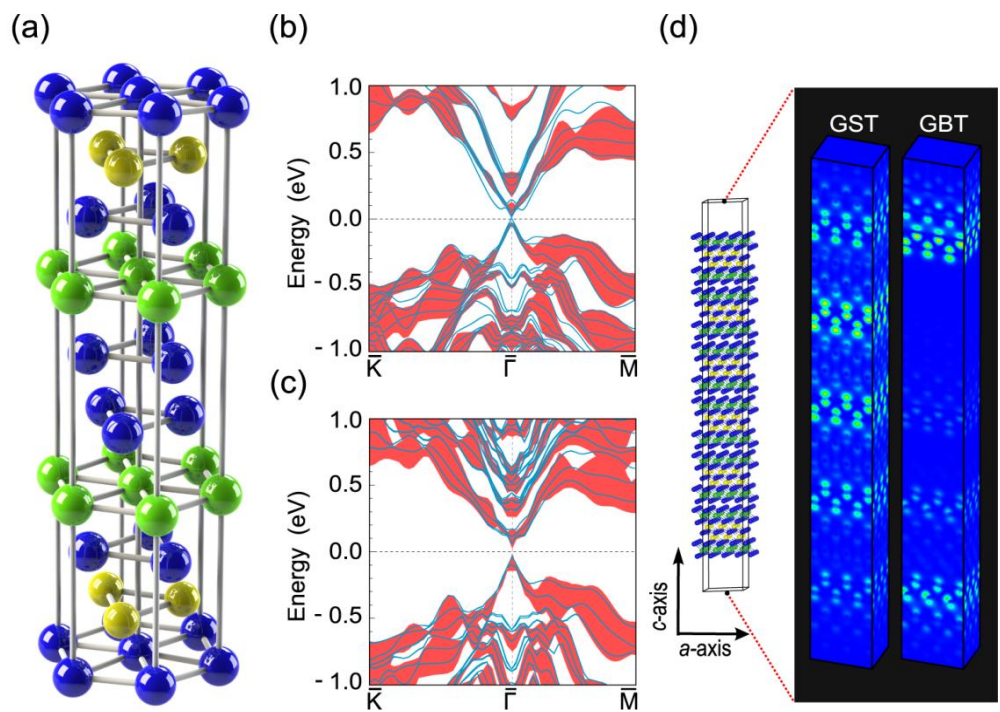

Figure 2 (Color online) Calculated energy gap at $\Gamma$ point for GST compounds in the order of $Sb_2Te_3$ layer thickness relative to GeTe layer thickness. In the top, the schematic band alignment is shown. The negative energy gap means the band inversion, which is a proof of the transition into the topological-insulating phase. We observe a transition of topological insulating phase (TP) – non-topological insulating phase (NTP) as the thickness is changed.

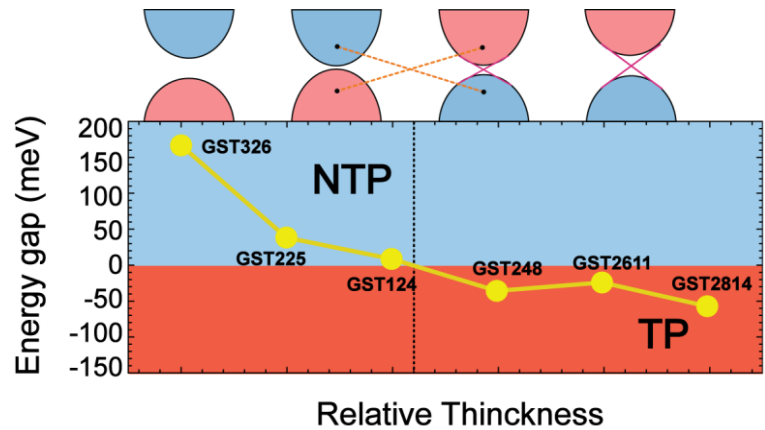

Figure 3 (Color online) (a) Schematic drawing of Dirac fermions in 1D superlattice. The surface states from stacked TI layers are represented by 1D array of Dirac cones on the right. (b) Energy diagram of TI-BI-TI junctions. Δ in the Dirac cone denotes the mass gap of TI layer due to the intralayer interaction of surface states and $\varepsilon_c$ is the chemical potential of band insulating layer of thickness *L* (half the energy gap with the Fermi level set at zero energy). (c) Schematic drawing of the helicity ordering of Dirac fermions in two adjacent TI layers (boxes in solid line). The momentum direction of the Dirac fermions is indicated by arrows and the spin direction by dot (out of paper) and x (into the paper) inside the circles. The upper panels represents the counter-helicity and the lower the co-helicity. Corresponding band dispersions (Dirac cones) are shown in the right with spin directions.

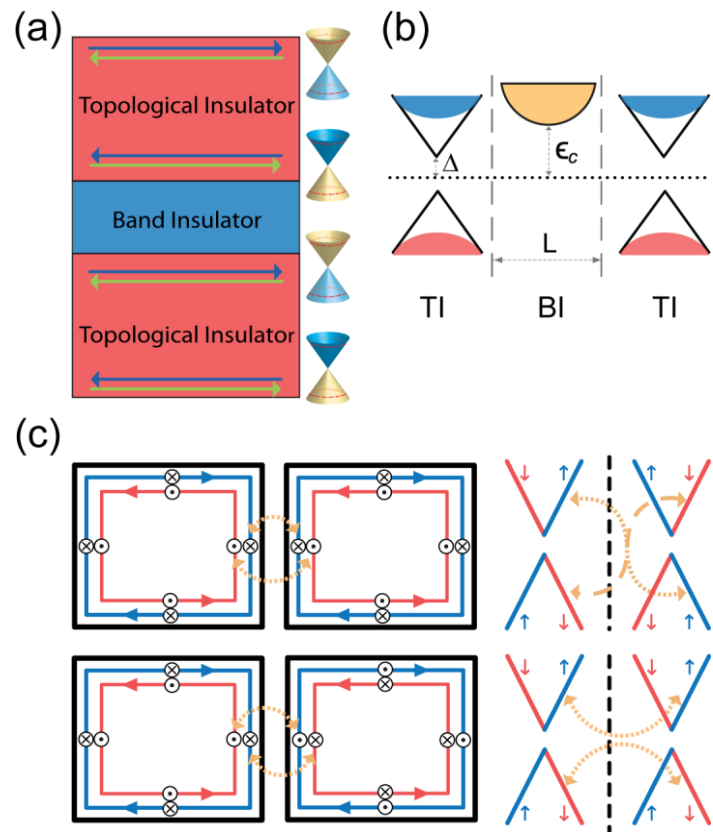

Figure 4 (Color online) (a) Topological insulating phase (TP) and non-topological insulating phase (NTP) of Dirac fermionic superlattice as a function of $\Delta$, $V$, and $\varepsilon_c$ from the superlattice model. (b) The projection of the phase diagram in (a) into 2D in terms of reduced parameters $\tilde{V}(=V/\varepsilon_c)$ and $\tilde{\Delta}(=\Delta/\varepsilon_c)$. Small dots are the numerical results of the effective Lagrangian and the solid line is a guide to the eyes. First-principles calculations of ternary compounds are also shown with the numbers denoting the composition of the compounds. (c), (d) Calculated helicity ordering of the surface states from TI layers in GST326 and GST248, respectively, using first-principles methods. The colored blocks represent the band insulating GeTe (blue) and topological insulating $Sb_2Te_3$ (red) layers. Arrows in TI layers represent the spin direction of the surface states in the $k_x$-$k_y$ plane of the momentum space. On the right is the simplified helicity of the surface states as a circular motion.

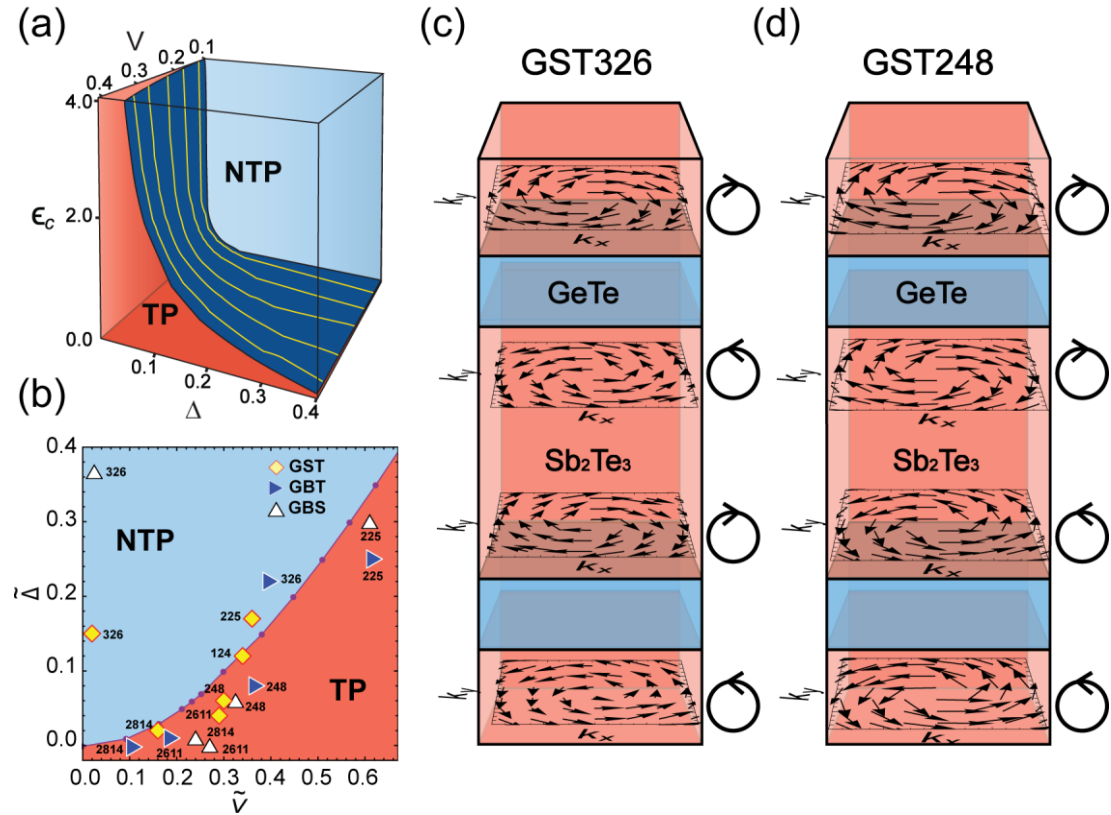